# Probing individual topological tunnelling events of a quantum field via their macroscopic consequences

Mitrabhanu Sahu[1], Myung-Ho Bae[1], Andrey Rogachev[1, 2], David Pekker[1, 3], Tzu-Chieh Wei[1, 4], Nayana Shah[1], Paul M. Goldbart[1] & Alexey Bezryadin[1]

[1]*Department of Physics, University of Illinois at Urbana Champaign*

[2]*Department of Physics, University of Utah*

[3]*Department of Physics, Harvard University*

[4]*Institute for Quantum Computing and Department of Physics and Astronomy, University of Waterloo*

**Phase slips are topological fluctuation events that carry the superconducting order-parameter field between distinct current carrying states[1]. Owing to these phase slips low-dimensional superconductors acquire electrical resistance[2]. In quasi-one-dimensional nanowires it is well known that at higher temperatures phase slips occur via the process of thermal barrier-crossing by the order-parameter field. At low temperatures, the general expectation is that phase slips should proceed via quantum tunnelling events, which are known as quantum phase slips (QPS). However, resistive measurements have produced evidence both pro[3-6] and con[7-9] and hence the precise requirements for the observation of QPS are yet to be established firmly. Here we report strong evidence for individual quantum tunnelling events undergone by the superconducting order-parameter field in homogeneous nanowires. We accomplish this via measurements of the distribution of switching currents—the high-bias currents at which superconductivity gives way to resistive behaviour—whose width exhibits a rather counter-intuitive, monotonic *increase* with decreasing temperature. We outline a**

**stochastic model of phase slip kinetics which relates the basic phase slip rates to switching rates[10,11]. Comparison with this model indicates that the phase predominantly slips via thermal activation at high temperatures but at sufficiently low temperatures switching is caused by individual topological tunnelling events of the order-parameter field, i.e., QPS. Importantly, measurements on several wires show that quantum fluctuations tend to dominate over thermal fluctuations at larger temperatures in wires having larger critical currents. This fact provides strong supports the view that the anomalously high switching rates observed at low temperatures are indeed due to QPS, and not consequences of extraneous noise or hidden inhomogeneity of the wire. In view of the QPS that they exhibit, superconducting nanowires are important candidates for qubit implementations[12,13].**

Quantum phenomena involving systems far larger than individual atoms are one of the most exciting fields of modern physics. Initiated by Leggett more than twenty-five years ago[14,15], the field has seen widespread development, important realizations being furnished, *e. g.*, by macroscopic quantum tunnelling (MQT) of the phase in Josephson junctions, and of the magnetization in magnetic nanoparticles[16-19]. More recently, the breakthrough recognition of the potential advantages of quantum-based computational methods has initiated the search for viable implementations of qubits[20], several of which are rooted in MQT in superconducting systems. In particular, it has been recently proposed that superconducting nanowires (SCNWs) could provide a valuable setting for realizing qubits[12]. In this case, the essential behaviour needed of SCNWs that they undergo QPS, i.e., topological quantum fluctuations of the superconducting order-parameter field via which tunnelling occurs between current-carrying states. It has also been proposed that QPS in nanowires could allow one to build a current standard, and thus could play a useful role in aspects of metrology[13]. Additionally, QPS are believed to provide the pivotal processes underpinning the superconductor-insulator transition observed in nanowires[21-25]. Observations of QPS have been reported previously on wires having high normal resistance (i.e., $R_N > R_Q$, where $R_Q = h/4e^2 \approx 6{,}450\ \Omega$) via low-bias resistance ($R$) vs. temperature ($T$) measurements[3,4]. Yet, low-bias measurements on short wires with normal resistance $R_N < R_Q$ have been unable to reveal QPS[7,8]. Also, it has been suggested that some results ascribed to QPS could in fact have originated in inhomogeneity of the nanowires. Thus, no consensus exists about the conditions under which QPS occur, and qualitatively new evidence for QPS remains highly desirable.

In this Letter, we present measurements of the distribution of stochastic switching currents—the high-bias currents at which the resistance exhibits a sharp jump from a very small value to a much larger one, close to $R_N$—in $Mo_{79}Ge_{21}$ nanowires. We observe a monotonic increase in the width of the distribution as the temperature is decreased. We analyze these findings in the light of a new theoretical model[11], which incorporates Joule-heating[10] caused by stochastically-occurring phase slips. The switching rates yielded by the model are quantitatively consistent with the data, over the entire range of temperatures at which measurements were performed (i.e., 0.3 K to 2.2 K), provided that both QPS and thermally-activated phase slip (TAPS) processes are included. By contrast, if only TAPSs are included, the model fails to give qualitative agreement with our observed switching-rate behaviour below 1.2 K. Thus, we conclude that in our SCNWs the phase of the superconducting order-parameter field slips predominantly via thermal activation at high temperatures; however, at temperatures below 1.2 K it is quantum tunnelling that dominates the phase-slip rate. It is especially noteworthy that at even lower temperatures (i.e., below 0.7 K) both our data and the model suggest that individual phase slips are, by themselves, capable of causing switching to the resistive state. Thus, in this regime, one has the capability of exploring the physics of single quantum phase-slip events. Furthermore, we observe strong effects of QPS at high bias currents, even in wires with $R_N < R_Q$. Another crucial fact is that the observed quantum behaviour is more pronounced in samples exhibiting larger switching currents. This fact allows us to rule out the possibility that the observed behaviour is caused by noise or wire inhomogeneity.

The linear low-bias resistance for the wire S1 is shown in Fig. 1a. The resistance measured just below the temperature at which the thin-film leads become superconducting is taken as the normal-state resistance $R_N$ of the wire. We find that the superconducting transition of the wire is well described by the phenomenological model of TAPSs in a quasi-1D superconductor developed by Langer-Ambegaokar and McCumber-Halperin (LAMH)[26,27]. To fit the $R$ vs. $T$ data, we have used the expression: $R(T) = \left[ R_{\text{LAMH}}^{-1}(T) + R_N^{-1} \right]^{-1}$ ([4]). The resistance due to TAPS is given by, $R_{LAMH}(T) = \dfrac{\pi \hbar^2 \Omega(T)}{2e^2 k_B T} \exp\left( -\dfrac{\Delta F(T)}{k_B T} \right)$, where $\Delta F(T)$ is the free-energy barrier for a phase slip in the zero-bias regime, and $\Omega(T)$ is the attempt frequency (Supplementary Information text)[2]. The fitting parameters used are the critical temperature $T_C = 4.34$ K, and the zero-temperature Ginzburg-Landau (GL) coherence length $\xi(0) = 8.2$ nm. As the temperature is decreased, the resistance becomes exponentially suppressed, and eventually falls below our experimental resolution. In contrast with the works of Giordano[3] and Lau $et$ $al.$[4], we do not find tails in the $R$ vs. $T$ data, which would indicate QPS. As an attempt to find the QPS regime we have opted to make measurements at high bias-currents, near the critical current, at low temperatures, i.e., in the regime in which the QPS rate should exceed the TAPS rate[10].

A representative set of voltage-current characteristics $V(I)$, measured at various temperatures, is shown in Fig. 1b These data show that, as the bias current is swept from low to high, the system exhibits an abrupt transition from a zero or extremely low-voltage state (i.e., a superconducting state) to a high-voltage state (i.e., a normal state) in which the resistance is close to $R_N$. We call the current at which switching occurs the switching current $I_{SW}$. Similarly, as the bias current is swept from high to low, the state reverts to being superconducting, doing so at a retrapping current $I_R$. We indicate these currents in Fig. 1b for data taken at 0.3 K. As can be seen from Fig. 1b, our nanowires are strongly hysteretic: there is a regime of currents within which the wire is bistable (i.e., two voltage states, one superconductive and one normal, are locally stable), and one of the two states is realized depending upon the history of the current sweep. We also find $I_{SW}$ is stochastic while $I_R$ is not, within our experimental resolution ($\sim 0.5$ nA).

We have observed that even when the temperature and current-sweep protocol are kept fixed, $I_{SW}$ varies from run to run, resulting in a distribution of switching currents $P(I_{SW})$, as was first studied for Josephson junctions by Fulton and Dunkleberger (FD)[28]. Such distributions, obtained at various temperatures, reflect the underlying, stochastically-fluctuating, collective dynamics of the condensate, and therefore provide a powerful tool for shedding light on the nature of the quasi-1D superconductivity. Indeed, one would expect the distribution width to scale with the thermal noise, and hence to decrease, as the temperature is reduced[28]; and to saturate at low temperature where thermal fluctuations are frozen out and only quantum fluctuations are left[18].

To obtain $P(I_{SW})$ at a particular temperature, we applied a triangular-wave current (sweep rate 125.5 µA/sec and amplitude 2.75 µA ), and recorded $I_{SW}$ (see Fig. 1b) for each of 10,000 cycles. We repeated this procedure at 21 equally spaced temperatures between 0.3 K and 2.3 K, thus arriving at the normalized distributions shown in Fig. 2. We observe the broadening of the switching-current distribution as the temperature is lowered, which is the exact opposite of the FD result[28]. This is our main observation, which is analyzed in detail below. This trend is confirmed by the analysis of the standard deviations σ of the distributions for samples S1, S2, S3 and S4; see Fig. 2 (inset). The ($R_N$, $L$) for these four samples, S1, S2, S3 and S4 are (2662 Ω, 110 nm), (4100 Ω, 195 nm), (1430 Ω, 104 nm) and (3900 Ω, 200 nm), respectively. We have transformed our $P(I_{SW})$ data into information on the rates $\Gamma_{SW}(I,T)$ at which switching would occur at a fixed current and temperature[28]. The switching rates resulting from the data in Fig. 2 are shown in Fig. 3.

To understand the origin of the peculiar dependence of the switching current distribution on temperature, we review mechanisms that could be responsible for the switching from the superconducting to the resistive state, and their implications for the switching current distributions. It is evident from the observed variability of the switching current that, to be viable, a candidate for the switching mechanism must be stochastic in nature. This suggests that the switching events are triggered by intrinsic fluctuations in the wire. In what follows, we shall focus on mechanisms driven by phase-slip fluctuations.

The simplest mechanism to consider is the one in which a single phase slip necessarily causes switching to the resistive state, as in an under-damped Josephson junctions[28]. In fact in our wires, at temperatures $T > \sim 1$ K, the rate of TAPS as indicated by both low-bias $R$-$T$ and high-bias $V$-$I$ measurements, is always expected to be much larger than the observed switching rate, even at very low currents. Therefore, at these temperatures, a current-carrying wire undergoes many TAPSs before the switch takes place, as directly confirmed by the non-zero voltage regime observed prior to the switching[8,9]; as shown in Fig. 1c (also see Fig. S1 and Supplementary Information text). For $T > 2.7$ K we can measure these residual voltage tails occurring at current lower than the switching current. As the temperature is reduced, these voltage tails, indicating a non-zero phase slips rate, become smaller, and below $\sim 2.5$ K the voltage falls below the experimental resolution of our setup ($\sim 2$ μV). The quantitative analysis of the switching process[11] leads us to the conclusion that the switching is activated by multiple phase slips at T $> \sim$ 1K and by single phase slips at T $< \sim$ 1K.

We now focus on switching mechanisms that incorporate multiple phase-slips. The observed high-voltage state is inconsistent with the presence of a phase-slip centre, because there is almost no offset current. We therefore hypothesize that the dynamics is always over-damped, and propose a runaway overheating model in spirit of ref. 10. Our model has two ingredients: (i) Stochastic phase slips that heat the wire by a quantum of energy $I\,h/2e$, and occur at random times and locations in the wire, but with a rate that depends on the local temperature of the wire. (ii) The heat produced by the phase slips is conducted along the wire, and is carried away by the leads. In effect, right after a phase-slip has occurred, the temperature of the wire rises, and therefore the phase-slip rate is enhanced. The higher phase-slip rate persists until the wire cools down. If another phase-slip happens to occur before the wire has cooled down, the temperature would rise further. Moreover, if, after several consecutive phase-slips, the temperature in the wire becomes high enough for the phase-slip rate to exceed the cooling rate, a subsequent cascade of phase slips carries the wire into the high-voltage state. Thus the switching is stochastic in nature. The rate of this switching is directly determined by the likelihood of having an initial burst of phase-slips that starts a cascade. This phenomenology is captured in Fig. 4a, which shows the temperature at the centre of a wire (above the lead temperature) as a function of time. Phase-slips correspond to sudden jumps in temperature, while cooling corresponds to the gradual decrease of temperature. A burst of phase-slips that results in a cascade can be seen near time t = 3 ns.

In the overheating model just discussed, the width of the switching-current distribution is controlled by the competition between the number of phase slips in the cascade-triggering burst and the rate of phase slips. If the number of phase slips to make such a burst tends to unity, the switching rate approaches the phase-slip rate. In the opposite regime, in which a large number of phase-slips are required to form the burst, the switching rate is much lower than the phase-slip rate. At higher temperatures, many phase slips are needed in the initial burst, and thus switching tends to occur only when $I_{SW}$ is in a very narrow range close to $I_C$, thus making the distribution narrow. As the temperature decreases, the heat capacity and heat conductivity both decrease, making phase-slips more effective at heating the wire. Thus, the typical number of phase slips in the cascade-triggering burst decreases with temperature, as our model shows[11]. At the same time, the rate of TAPS also decreases with temperature. In practice, with decreasing temperature, the broadening effect of the burst length on $P_{SW}(I)$ overwhelms the narrowing effect of the decreasing TAPS rate, and this provides a possible explanation of the unanticipated broadening of the $I_{SW}$ distributions.

We first tried to fit all the switching-rate data in Fig. 3 using the overheating model but with a phase-slip rate $\Gamma$ that follows from allowing only thermally-activated (and not quantum) processes, i.e., $\Gamma_{TAPS}$. At temperature $T$ and bias-current $I$, $\Gamma_{TAPS}$ is given by,

$$\Gamma_{TAPS}(T, I) = \frac{\Omega_{TAPS}(T)}{2\pi} \exp\left(-\frac{\Delta F(T, I)}{k_B T}\right) \qquad (1)$$

where $\Omega_{TAPS}(T)$ is the attempt frequency, $\Delta F(T, I) = \Delta F(T)\left(1 - \left(\frac{I}{I_C(T)}\right)\right)^{5/4}$ is the free-energy barrier at bias-current $I$ ([10,26]), $I_C(T)$ is the fluctuation-free de-pairing current, and $\Delta F(T) = \sqrt{6}\frac{\hbar I_C(T)}{2e}$ is the free-energy barrier at zero bias-current[29] (Supplementary Information text). These fits agree well with the data over the temperature range 2.4 K to 1.3 K (Fig. 3a). Within this range, we can attribute the decrease in the width of the distribution to the mechanism described in the previous paragraph: the competition between (i) the number of phase slips in the initial burst required to start a cascade, and (ii) the rate of phase slips. However, below 1.2 K it is evident from Fig. 3a that the switching rates predicted by TAPS are considerably smaller than the switching rates obtained experimentally.

As the temperature is reduced and fluctuations become smaller, the switching happens at higher values of the bias-current $I$. Thus each phase slip releases more heat into the wire ($I h/2e$). Also, as bias-current $I$ is increased, the value of the temperature increase required in order to reach the normal state becomes smaller. Therefore, according to the overheating model one ultimately expects to have a low $T$ regime in which a single phase-slip event releases enough heat to induce a switching event[11]. We call this the single-slip regime. We expect that for $T < \sim 0.7$ K, our wires should be operating in this single-slip regime, as indicated in Fig. 3b[11]. We find, however, in the regime 0.3 K $< T <$ 1.2 K our data can not be fitted well if the phase-slip rate is taken to be $\Gamma_{\text{TAPS}}$, but can be fitted well if the total phase-slip rate ($\Gamma_{\text{TOTAL}}$) is taken to be the sum of TAPS rate ($\Gamma_{\text{TAPS}}$) and QPS rate ($\Gamma_{\text{QPS}}$) (i.e., $\Gamma_{\text{TOTAL}} = \Gamma_{\text{TAPS}} + \Gamma_{\text{QPS}}$). Since at 0.3 K we are already in the single-slip regime the switching rate should be equal to the phase-slip rate. As shown in Fig. 4b, at 0.3 K, the measured switching rate can be fitted by the Giordano-type QPS rate[3], given by the same expression as the TAPS rate but with the wire temperature $T$ replaced by an effective "quantum" temperature $T_{\text{QPS}}(T)$, i.e.,

$$\Gamma_{\text{QPS}}(T, I) = \frac{\Omega_{\text{QPS}}(T)}{2\pi} \exp\left(-\frac{\Delta F(T, I)}{k_{\text{B}} T_{\text{QPS}}}\right) \text{ (3,4)}$$

For $T = 0.3$ K, the switching rate predicted by TAPS is roughly $10^{15}$ times smaller than the measured switching rate (Fig. 4b top panel). Using different expressions for the attempt frequency (e.g. those derived for Josephson junctions) can only increase the disagreement between the TAPS model and the data (see Supplementary Information text and Fig. S3). On the other hand, fitting the measured switching rate with the Giordano-type QPS rate for several values of the temperature; we find a very good agreement. The corresponding effective quantum temperature $T_{\text{QPS}}(T)$ is considerably higher than the bath temperature $T$, which is a strong indication of QPS. We also observe that to fit data it is necessary to assume a weak linear dependence of the $T_{\text{QPS}}(T)$ on the bath temperature $T$ (Fig. 4c). For sample S1, $T_{\text{QPS}}$ is found to be, $T_{\text{QPS}}(T) = 0.726 + 0.40 \times T$ (in Kelvins). The non-zero intercept indicates the persistence of the high-bias-current-induced QPS down to zero temperature. It is found that below a crossover temperature $T^*$ the QPS rate dominates over the TAPS rate and the fluctuations in nanowire are mostly quantum in nature. This $T^*$ for wire S1 is 1.2 K and is denoted by the red arrow in Fig. 4c (see Supplementary Information text). To verify the consistency of our model at all measured temperatures, we replaced the TAPS rate by the total phase-slip rate $\Gamma_{\text{TOTAL}}$ to obtain the switching rates over the full range of temperatures (i.e., 0.3 K-2.3 K). We find that the predicted switching rates agree reasonably well with the data as shown in Fig. 3b for all temperatures.

Furthermore, we verified the evidence of QPS in three more nanowire samples (S2-S4). The linear dependence of $T_{\text{QPS}}(T)$ for these nanowires is shown in Fig. 4c. As with the first sample, this linear dependence is chosen to give the best possible fits to the measured switching rates, as those in Fig. 4b. Also, the corresponding crossover temperatures $T^*$ for all samples are indicated by the arrows. We find that the $T^*$ is consistently reduced with the reduction of the critical depairing current $I_{\text{C}}(0)$; as shown in Fig. 4c (see Supplementary Information for details). This observed trend is analogous to the case of Josephson junctions in ref. 17. To understand this observation we remind that $T^*$ is proportional to the plasma frequency of the device, which, in turn, is proportional to the critical current. On the other hand, if the observed increase in the fluctuation strength and the fact that $T_{\text{QPS}} > T$ were due to some uncontrolled external noise in the setup, the thicker wires, having larger critical currents, would have shown a reduction in the $T^*$, which is opposite to what we observe. These observations also allow us to rule out the possibility that some hidden granularity cause the QPS-like effects. Indeed, what we find is that wires of lower critical currents, which obviously have more chance to have weak links, show a less pronounced quantum behaviour and a lower $T^*$ value (Fig.4c). Thus the possibility of weak links producing the reported here QPS-like effects is ruled out. In conclusion, the result of Fig.4c provides a qualitatively new and strong evidence for the existence of QPS in thin superconducting wires.

In conclusion, we have measured distributions of the currents at which switching from the superconducting state to the normal state occurs, for a range of temperatures. We have also applied a stochastic model that relates phase-slip rates to the switching rates and explained the broadening of the switching-current distributions with cooling, observed at higher temperatures. In addition, we have been able to identify and explore a low-temperature regime in which switching events are triggered by single phase slips, as the model shows. Thus, we have been able to extract temperature-dependant rates for phase-slip processes. These extracted rates strongly suggest that at lower temperatures, we are observing the consequence of the quantum tunnelling of a macroscopic variable, namely, the superconducting order-parameter field, i.e., quantum phase slips. As expected for QPS, we observe that they start to dominate TAPS at higher temperatures in wires of higher critical current. This observation provides strong evidence that the observed high rate of switching at low temperatures is indeed due to quantum tunnelling, and not to extraneous noise or wire inhomogeneity.

**Methods**

Our nanowires were fabricated using molecular templating[21]. Amorphous $Mo_{79}Ge_{21}$ alloy was sputtered onto fluorinated single-wall carbon nanotubes that were suspended across 100-200 nm wide trenches[30]. The wires appear quite homogeneous on SEM images, such as in the inset of Fig. 1a. The nanowire is seamlessly connected to thin-film $Mo_{79}Ge_{21}$ leads at each of its ends. All of our measurements were performed in $^3$He cryostat with the base temperature ~ 285 mK. All signal lines were equipped with room-temperature (7 dB cut-off frequency of 3 MHz, Spectrum Control) and cooper powder and silver-paste microwave filters kept at the base temperature (see Supplementary Information). For the signal lines with all the filtering the measured attenuation is larger than 100 dB for frequencies higher than 1 GHz (Fig. S6). The voltage signals were amplified using battery-powered, low-noise preamplifiers (SR 560). The samples were measured with a four-probe configuration as described in ref. 21.

---

**Acknowledgements** This material is based upon work supported by the U.S. Department of Energy, Division of Materials Sciences under Award No. DE-FG02-07ER46453, through the Frederick Seitz

Materials Research Laboratory at the University of Illinois at Urbana-Champaign. One of us (M.-H.B.) acknowledges the support of the Korea Research Foundation Grants No. KRF-2006-352-C00020.

**Author Contributions** M.S. fabricated all the nanowire samples; M.S., M.-H.B., A.R. and A.B. performed all the measurements; M.S., M.-H.B., D.P., N.S, T.-C.W., P.G. and A.B. worked on the theoretical modelling, data analysis and co-wrote the paper. All authors discussed the results and commented on the manuscript.

**Author Information** Reprints and permissions information is available at npg.nature.com/reprints .Correspondence and requests for materials should be addressed to M.S. ([sahu@illinois.edu](mailto:sahu@illinois.edu)).

**Figure 1 | Basic sample characterization at high and low temperatures. a,**
Zero-bias resistance vs. temperature measurement (sample S1). The circles
are the data and the solid line is the fit to the LAMH model. The normal-state
resistance of the wire, $R_N$ is indicated by the arrow, which is measured
immediately below the film transition. Our setup allows measurements down to
0.1 Ω, which is reached already at T=3.0 K. (Inset) Scanning electron
microscope (SEM) image of nanowire sample S1. **b,** Voltage vs. current
dependence at various temperatures from 2.3 K to 0.3 K for wire S1. The
switching current $I_{SW}$ and retrapping current $I_R$ are indicated for the data taken
at 0.3 K. The temperatures are T = 0.3 (the highest $I_{SW}$), 0.6, 0.9, 1.2, 1.5, 1.8,
2.1, 2.4 K. **c,** Residual voltage tails observed at high-bias currents, just before
the switching. The temperatures are $T$ = 3.44, 3.33, 3.11, 3.0, 2.85, 2.49,
2.29 K. These voltage tails becomes smaller as the temperature is decreased
and becomes immeasurably low for $T$ < ~ 2.5 K.

**Figure 2 | Switching current distributions at different temperatures.**
Switching current distributions $P(I_{SW})$ for temperatures between 0.3 K (right
most) and 2.3 K (left most) with $\Delta T$ =0.1 K for sample S1. For each distribution
10,000 switching events were recorded and the bin size of the histograms was
3 nA. **inset,** Standard deviation σ $\left( = \sqrt{\left( \sum_{i=1}^{n} (I_{SW,i} - \bar{I}_{SW})^2 \right)/(n-1)} \right)$ of $P(I_{SW})$ vs. $T$
for four different nanowires including sample S1. For samples S1 and S2 the
measurements were repeated a few times to verify the reproducibility of the
temperature dependence of σ. For all wires, the width of the distributions
increases as the temperature is decreased.

**Figure 3 | Measured switching rates from the superconducting state and predictions of the stochastic overheating model. a,** Switching rates from the superconducting state to the resistive state for bath temperatures between 2.3 K (left most) and 0.7 K (right most) (here not all the measured curves are shown for clarity). The data is shown for all temperatures between 2.3 K and 1.1 K with $\Delta T = 0.1$ K and for $T = 0.9$ K and $T = 0.7$ K (sample S1). The symbols are experimental data and the lines (with corresponding colour) are fits to the overheating model that incorporates stochastic TAPS-only (see text). The fits agree well with the data down to 1.3 K, indicated by an arrow. **b,** Fits to the same data (all temperatures are shown here) with the stochastic overheating model which now incorporates both the TAPS and QPS rates to calculate the switching rates. The boundary for single phase-slip switch regime is indicated by the black diamond symbols at four temperatures (connected by line segments). For the measured range of switching rates, any of the ($I$, $\Gamma_{SW}$) to the right of this boundary (i.e., for higher bias currents) is in the single-slip regime.

**Figure 4 | Stochastic phase slips, switching rates, and the quantum behaviour at low temperatures. a,** Simulated "temperature bumps" in the nanowire due to a sequence of phase-slips events. The bath temperature is assumed to be $T_b = 2.4$ K, $T_C = 3.87$ K and bias current $I = 1.0$ µA. As the temperature of the wire section becomes higher than $T_C$, it becomes normal. **b,** (top panel) Switching rates at $T = 0.3$ K for sample S1 (open circles). The blue curve is the fit to the data, based on the Giordano-type QPS model. The red curve is expected for the TAPS rate. The arrow indicates the difference between the expected TAPS rate and the data. This difference is very large namely $10^{15}$ Hz. (bottom panel) The corresponding switching current

distribution at 0.3 K (open circles) and the predictions due to QPS rate (blue) and TAPS rate (red).**c,** The best fit effective temperature for fluctuations at different bath temperatures for four different samples (S1-S4). For all TAPS rates calculation the effective temperature is chosen as the bath temperature (shown by the black dotted line). For the QPS rates, the effective temperature $T_{QPS}$, used in the corresponding QPS fits, similar to the blue-line fits of Fig.4b, are shown by the solid lines. For each sample, below the crossover temperature $T^*$ (indicated by arrows), QPS dominates the TAPS. We find that the $T^*$ decreases with decreasing critical depairing current of the nanowires, which is the strongest proof of QPS. The trend indicates that the observed behaviour of $T_{QPS}$ below $T^*$ is not due to extraneous noise in the setup or granularity of wires, but, indeed, is due to QPS.

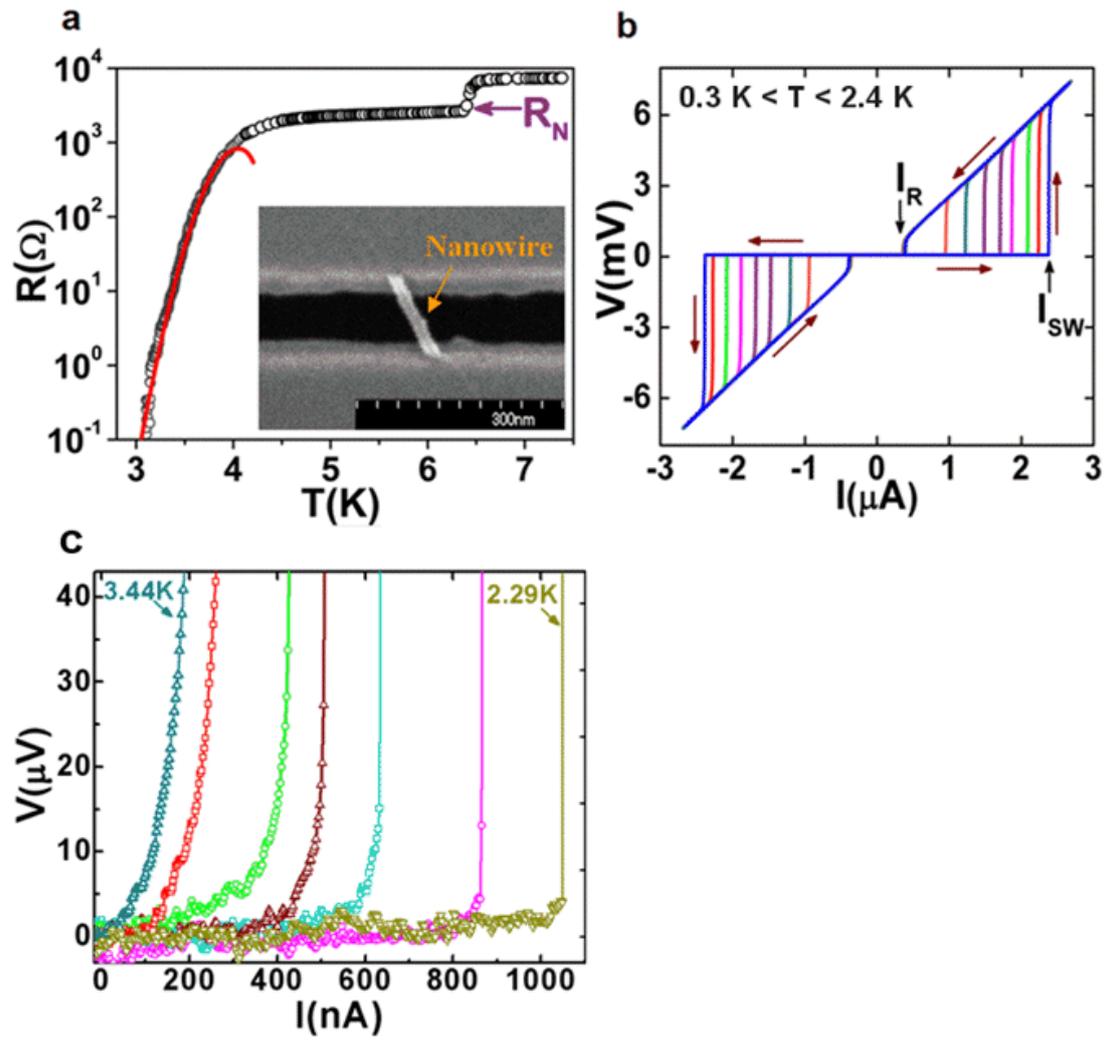

**Figure 1**

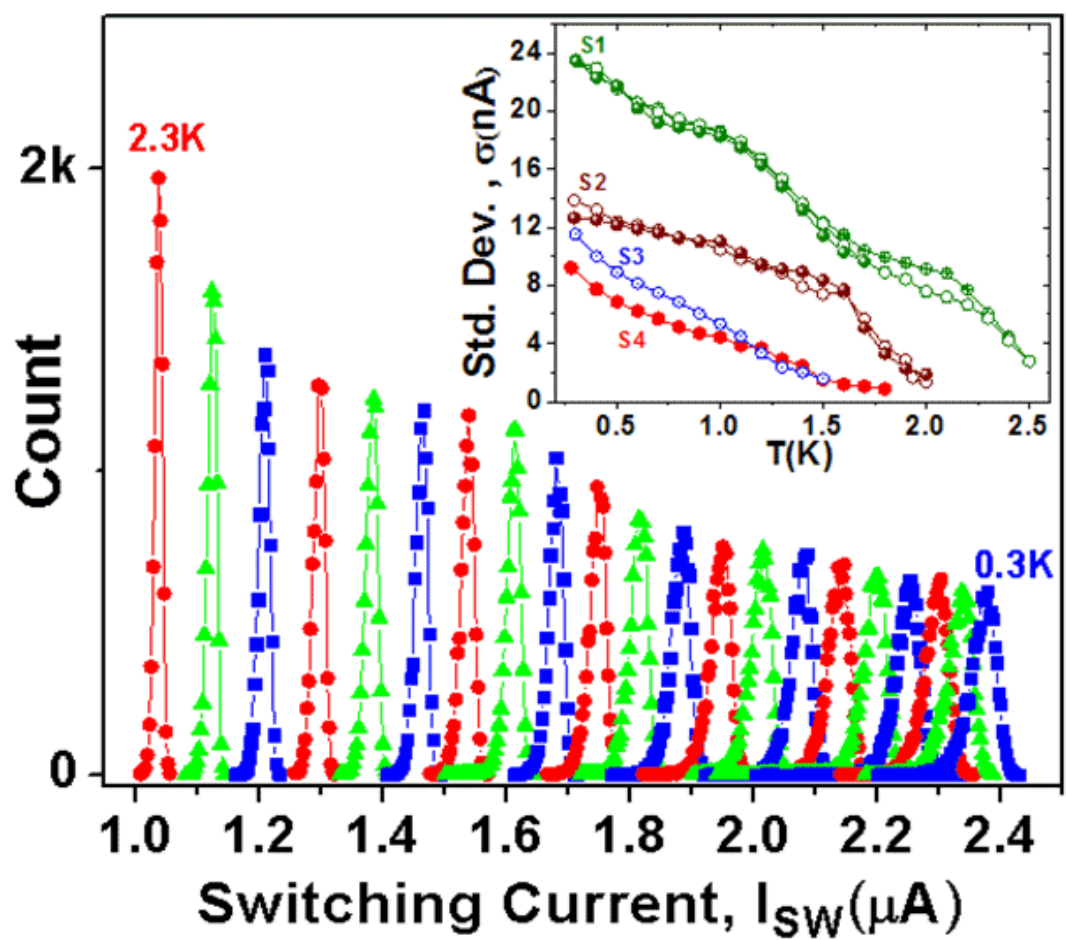

**Figure 2**

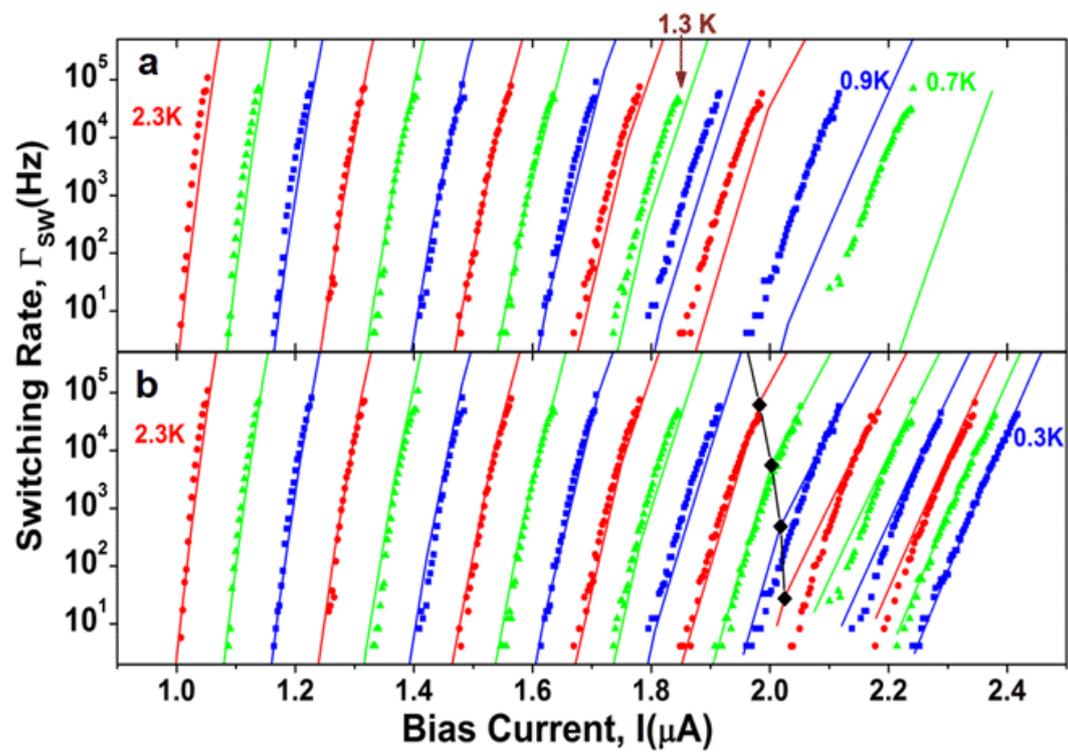

**Figure 3**

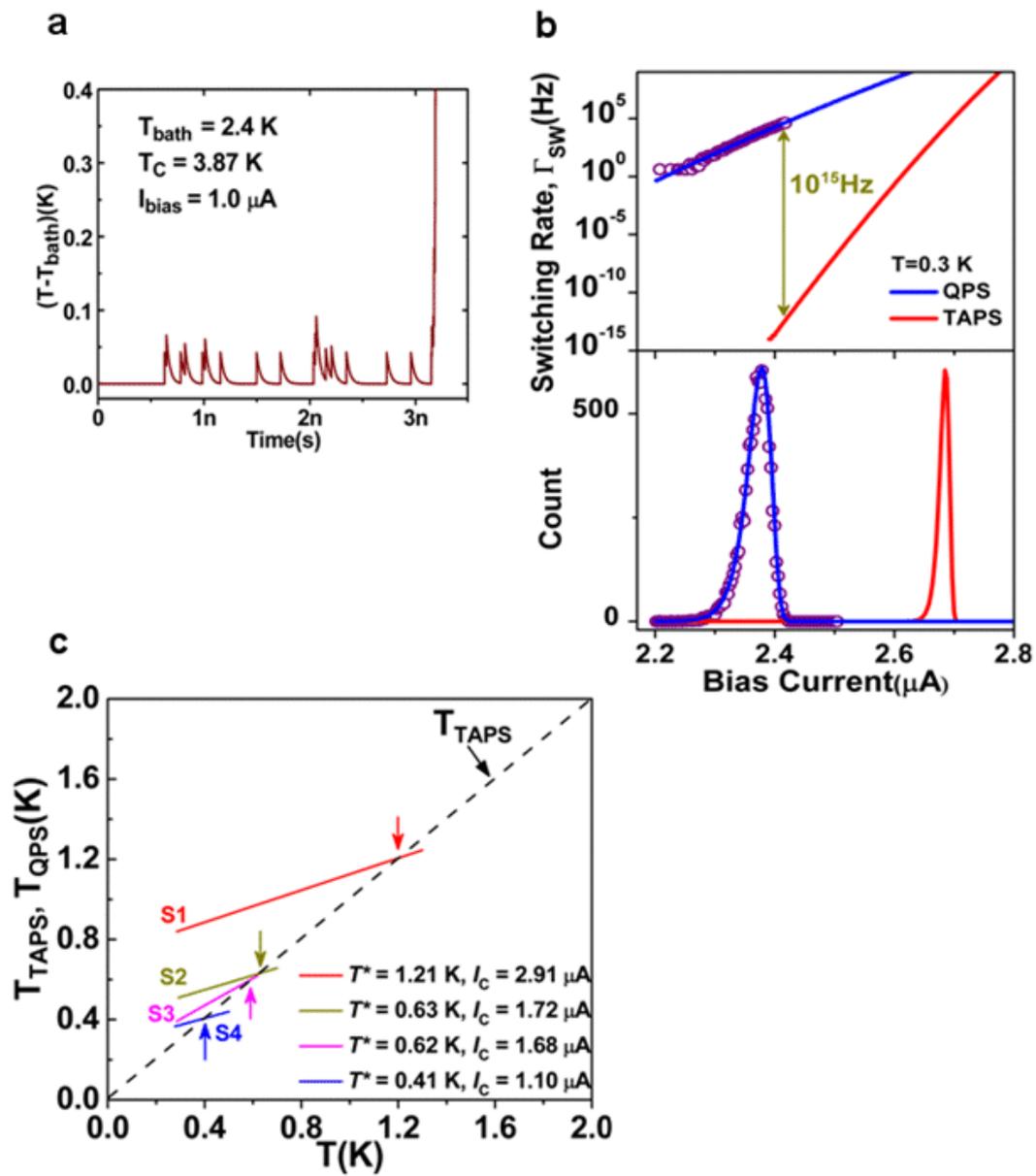

**Figure 4**

# Supplementary Information

## Probing individual topological tunnelling events of a quantum field via their macroscopic consequences

Mitrabhanu Sahu[1], Myung-Ho Bae[1], Andrey Rogachev[1,2], David Pekker[1,3],

Tzu-Chieh Wei[1,4], Nayana Shah[1], Paul M. Goldbart[1] & Alexey Bezryadin[1]

[1]*Department of Physics, University of Illinois at Urbana Champaign*

[2]*Department of Physics, University of Utah*

[3]*Department of Physics, Harvard University*

[4]*Institute for Quantum Computing and Department of Physics and Astronomy,
University of Waterloo*

**This section contains:**

1. **Discussion of the thermally activated phase slip (TAPS) and quantum phase slip (QPS) rates.**

2. **Macroscopic quantum tunnelling in high-$T_C$ intrinsic Josephson junctions.**

3. **Filtering system in our measurement setup.**

4. **Supplementary references.**

# **Discussion of the thermally activated phase slip (TAPS) and quantum phase slip (QPS) rates**

To fit the low-bias $R$ vs. $T$ data we used the expression,

$$R(T) = \left[ R_{\text{LAMH}}^{-1}(T) + R_{\text{N}}^{-1} \right]^{-1} \qquad (1)$$

Here, we have taken into account the normal conductance channel, which is due to quasi-particles and this conductance is typically estimated as $1/R_{\text{N}}$. This normal channel is connected in parallel with the conductance of the condensate in the wire ($R_{\text{LAMH}}$), which is not infinite due to TAPS ([1],[2]). The theory of TAPS, developed by Langer-Ambegaokar and McCumber-Halperin, is called LAMH ([3],[4]). According to this theory the resistance due to TAPS is given by

$$R_{LAMH}(T) = \frac{\pi \hbar^2 \Omega(T)}{2e^2 k_B T} \exp\left( -\frac{\Delta F(T)}{k_B T} \right) \qquad (2)$$

where $\Delta F(T) = \dfrac{8\sqrt{2}}{3} \left( \dfrac{H_{\text{C}}^2(T)}{8\pi} \right) A\xi(T)$ is the energy barrier for phase slips,

$\Omega(T) = \left( L/\xi(T) \right)\left( \Delta F(T)/k_B T \right)^{1/2} \left( 1/\tau_{GL} \right)$ is the attempt frequency, $\tau_{\text{GL}} = [\pi\hbar/8k_{\text{B}}(T_{\text{C}} - T)]$ is the Ginzburg-Landau (GL) relaxation time, $L$ is the length of the wire, $A$ is the cross-sectional area, $\xi(T)$ is the GL coherence length, $H_{\text{C}}(T)$ is the critical field and $T_{\text{C}}$ is the critical temperature of the wire. The $T$ dependence of the of the energy barrier $\Delta F(T)$ and the attempt frequency $\Omega(T)$ come in the expression via $\xi(T)$ and $H_{\text{C}}(T)$, which are given as

$$\xi(T) = \xi(0) \frac{\left( 1 - \left( \dfrac{T}{T_{\text{C}}} \right)^4 \right)^{0.5}}{\left( 1 - \left( \dfrac{T}{T_{\text{C}}} \right)^2 \right)} \qquad (3)$$

$$H_C(T) = H_C(0)\left[1.73\left(1-\frac{T}{T_C}\right) - 0.40087\left(1-\frac{T}{T_C}\right)^2 - 0.33844\left(1-\frac{T}{T_C}\right)^3 + 0.00722\left(1-\frac{T}{T_C}\right)^4\right] \quad (4)$$

Here we have found the temperature dependence of $H_C(T)$ by fitting the numerical tabulation given by Muhlschlegel to a polynomial fit (*2,[5]*) applicable at all temperatures below $T_C$. Also, the energy barrier at zero temperature $\Delta F(0)$ can be expressed in terms of wire parameters ([6]),

$$\Delta F(0) = \frac{1.76\sqrt{2}}{3}\left(\frac{R_Q}{R_N}\right)\left(\frac{L}{\xi(0)}\right)k_B T_C \quad (5)$$

where $R_Q = h/4e^2 \approx 6.45\text{k}\Omega$. The fitting parameters are $T_C$ and $\xi(0)$. The length $L$ of the wire is determined from SEM images. The normal resistance $R_N$ of the wire is taken to be resistance measured as the film electrodes, connected in series with the wire, become superconducting.

Alternatively, one can express the free energy barrier in terms of the critical depairing current $I_C(T)$ (*6,[7],[8],[9]*) as, $\Delta F(T) = \frac{\sqrt{6}\hbar I_C(T)}{2e}$, where,

$$I_C(T) = (92uA)\frac{LT_C}{R_N\xi(0)}\left(1-\left(\frac{T}{T_C}\right)^2\right)^{3/2}$$ (where, $L$ and $\xi(0)$ are in nm, $T_C$ is in K and $R_N$ is in $\Omega$). A more useful expression directly applicable for our high-bias measurements data, which takes into account both the temperature and bias-current dependence of the energy barrier $\Delta F(T,I)$, is given by (*7,8*),

$$\Delta F(T,I) = \frac{\sqrt{6}\hbar I_C(T)}{2e}\left(1-\frac{I}{I_C}\right)^{5/4} \quad (6)$$

The TAPS rate, $\Gamma_{TAPS}$ used in the overheating model is given by,

$$\Gamma_{\text{TAPS}} = \Omega_{\text{TAPS}} \exp\left(-\frac{\Delta F(T,I)}{k_{\text{B}}T}\right)$$

$$= \left(\frac{L}{\xi(T)}\right)\left(\frac{1}{\tau_{\text{GL}}}\right)\left(\frac{\Delta F(T)}{k_{\text{B}}T}\right)^{1/2} \exp\left(-\frac{\Delta F(T,I)}{k_{\text{B}}T}\right) \qquad (7)$$

A simple model of quantum phase slips was suggested by Giordano ([10]). We use this model, but instead of the Ginzburg-Landau relaxation time, which is only correct near $T_{\text{C}}$, we use the notion of the effective "quantum" temperature $T_{\text{QPS}}$, which is a common (and well-tested) approach in Josephson junctions (JJ) ([11]). Thus, the QPS rate, $\Gamma_{\text{QPS}}$ is given by,

$$\Gamma_{\text{QPS}} = \Omega_{\text{QPS}} \exp\left(-\frac{\Delta F(T,I)}{k_{\text{B}}T_{\text{QPS}}}\right)$$

$$= \left(\frac{L}{\xi(T)}\right)\left(\frac{1}{\tau_{\text{GL}}}\right)\left(\frac{\Delta F(T)}{k_{\text{B}}T_{\text{QPS}}}\right)^{1/2} \exp\left(-\frac{\Delta F(T,I)}{k_{\text{B}}T_{\text{QPS}}}\right) \qquad (8)$$

Here, $T_{\text{QPS}}$ is the effective quantum temperature representing the strength of zero-point fluctuations in the $LC$-circuit formed by the nanowire, which has a nonzero kinetic inductance of the order of 0.1 nH, and the leads, with a mutual capacitance of the order of 1-10 fF. Thus we can roughly estimate the plasma frequency as $1/\sqrt{LC}$ and therefore the expected quantum temperature is $\hbar/2\pi k_B\sqrt{LC} \sim 1$ K. Experimentally we indeed find that the quantum temperature is of the order of 1 K. We also note that that to obtain a good agreement between the experimental switching rate and the quantum model a week linear dependence of $T_{\text{QPS}}$ on the bath temperature $T$ has to be assumed (see Fig. 4c). More precisely, we use $T_{\text{QPS}}(T) = 0.726 + 0.40 \times T$ (in Kelvins) for sample S1. For all the TAPS and QPS rates the wire parameters [i.e., $R_{\text{N}}$, $L$, $T_{\text{C}}$, and $\xi(0)$] are kept the same. For example, for sample S1, $R_N$ = 2666 Ω, $L$ = 110 nm, $T_C$ = 3.872 K and ξ(0) = 5.038 nm.

Furthermore, we find that below a crossover temperature $T^*$, the QPS rate dominates over the TAPS rate, i.e., quantum fluctuations dominate over the thermally induced fluctuations. For sample S1, $T^*$ = 1.21 K (see Fig. 4c). Similar analysis on another three nanowires (S2-S4) reveals that the crossover temperature decreases with decreasing critical depairing current; as shown in Fig. 4c. This is an important fact since it leads to a conclusion that the observed QPS effect, i.e., the observation of $T_{QPS}$ to be higher than the bath temperature, is not due to some noise or weak links in the wires. This analysis is analogous to the discussion of macroscopic quantum tunnelling in JJ (see ref. 2, Fig.7.4 (page 263) in the paragraph about MQT). This increase of the $T^*$ with the critical current indicates that the observed large value of the width of the switching current distributions is an intrinsic property of the nanowires, occurring due to QPS. In the following table, we enlist the wire parameters that were used for all the four samples to get $\Gamma_{TAPS}$ and $\Gamma_{QPS}$ and their form of $T_{QPS}(T)$.

**Table 1| Nanowire sample parameters, $T_{QPS}(T)$ and $T^*$ for all samples**

| Nanowire Sample | L (nm) | $R_N$ (Ω) | $T_C$ (K) | ξ(0) (nm) | $I_C(0)$ (nA) | $T_{QPS}(T)$ (In the form a+b$T$) (K) | $T^*$ (K) |
|---|---|---|---|---|---|---|---|
| S1 | 110 | 2666 | 3.872 | 5.038 | 2917 | 0.726+0.40$T$ | 1.210 |
| S2 | 195 | 4100 | 3.810 | 9.650 | 1727 | 0.404+0.362$T$ | 0.633 |
| S3 | 104 | 1430 | 3.160 | 12.560 | 1683 | 0.199+0.678$T$ | 0.620 |
| S4 | 200 | 3900 | 2.870 | 12.250 | 3900 | 0.275+0.33$T$ | 0.410 |

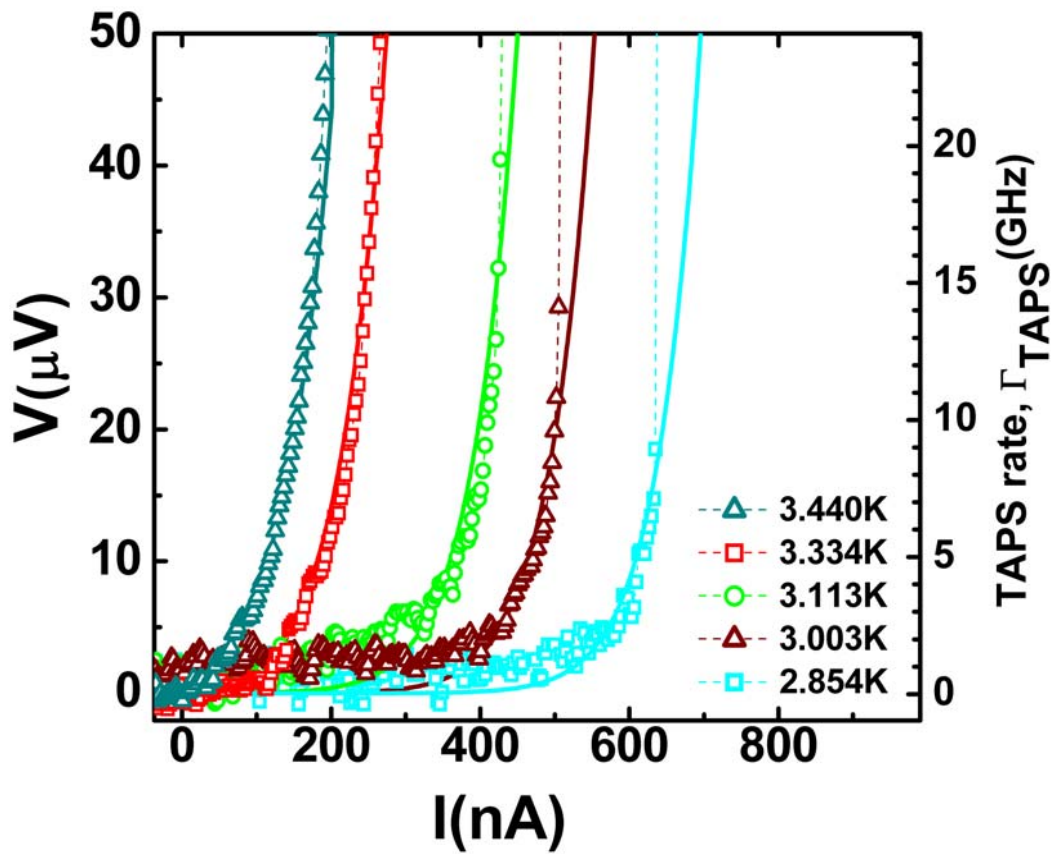

**FIG. S1:** High-bias V-I measurements at high temperatures where the voltage due to phase diffusion is measurable even before the switching event. The solid lines are predictions of phase slip rate using the TAPS model with the wire parameters used are those which were obtained by fitting the switching rates measurements between T = 0.3 K to 2.3 K. The phase-slip rate $\Gamma_{TAPS}$ (shown in the right axis) is converted into voltage using the relation, $V = \dfrac{h\Gamma_{TAPS}}{2e}$. The dashed line is the line connecting the data points, not a fit.

Another independent validation of the TAPS model, applied at higher temperatures, and the wire parameters used, can be obtained from non-linear *I-V* curves measured at relatively high temperatures. At these temperatures measured *I-V* curves show tails due to TAPS that are large enough to be measured in our set up; as shown in Fig. S1 (see also Fig. 1c). In Fig. S1, we also plotted the predicted voltage obtained from the TAPS model, using $V = h\Gamma_{TAPS} / 2e$. For the TAPS rate calculations, the wire parameters used for all our fittings (as shown in Fig. 3) are kept the same and only the $T$ was varied to get the corresponding TAPS rate. The measured voltage (or phase-slip rate) and the predicted voltage (or the TAPS rate) are in good agreement for the five temperatures noted in Fig. S1. This agreement indeed verifies our model for TAPS. The calculation of the TAPS $V\text{-}I$ curve is made under the assumption that the wire temperature equals the bath temperature, i.e., no significant Joule heating occurs. These type of phase diffusion "tails" on the $V\text{-}I$ curves can only be seen at temperature of about 2.7 K or larger, which is about 10 times higher a temperature than those where the QPS effects are found.

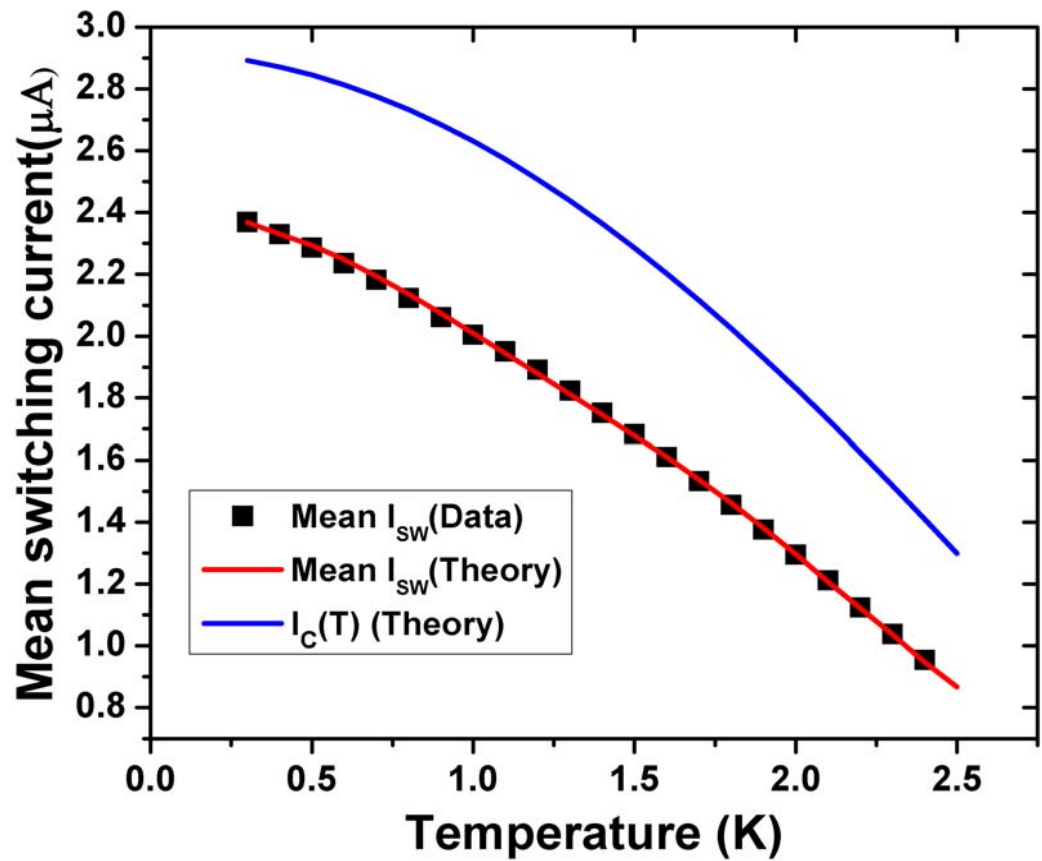

**FIG. S2:** The measured mean switching current (squares) and the predicted mean switching current by our model (red line) as a function of temperature.

The predicted fluctuation-free critical depairing current, $I_c(T)$ is shown (blue line). At all temperatures a premature switching occurs before the bias current reaches the critical depairing current.

The mean switching current predicted at each temperature by the overheating model and the mean switching current for each distribution (Mean $I_{SW}$) is compared in Fig. S2. We have also plotted the critical depairing current

$$I_C(T) = I_C(0)\left[1-\left(T/T_C\right)^2\right]^{3/2} (9).$$ We find that at all temperatures the switching is premature.

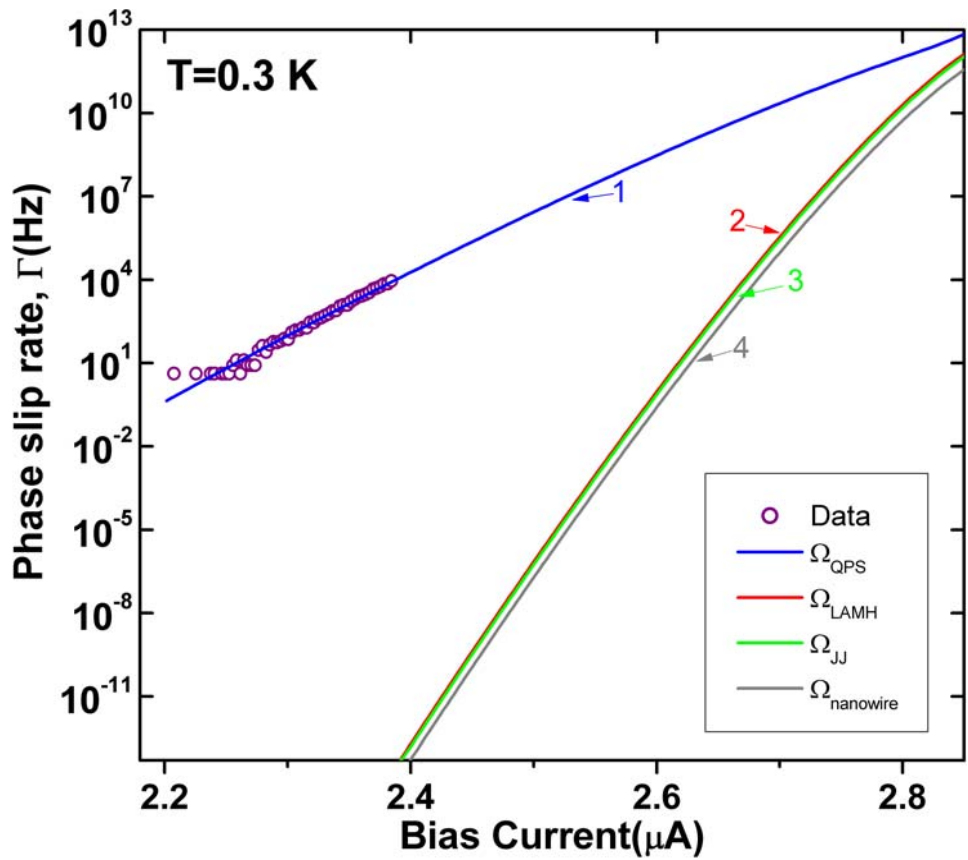

**FIG. S3:** The data (open circles) and the calculated QPS rate (solid blue line) at 0.3 K for wire S1. The observed agreement is very good. Different estimates of TAPS rate by using different attempt frequency expressions are also shown by solid red, green, and grey lines. For all our estimates of TAPS rate, the data is at least ~ $10^{15}$ orders of magnitude higher than the predicted thermal rate. Hence, the data can not be explained by considering thermal fluctuation alone, even if the uncertainty in the attempt frequency is taken into account. Note also that for the lowest bias current of 2.2 μA, the thermal rate is about $10^{25}$ orders lower than the experimental rate, which further proves our point.

We would like to briefly comment on the attempt frequency $\Omega$, that is used to get the TAPS rate. In Fig. S3, we have plotted the TAPS rates estimated using different expressions for $\Omega$ (curves 2 - 4), the data (open circles) and the QPS rate (curve 1) (all at 0.3 K). For curve 2 we have used, $\Omega = \left( L / \xi(T) \right) \left( \Delta F / k_B T \right)^{1/2} \left( 1 / \tau_{GL} \right)$ according to McCumber and Halperin expression (eq. 7), based on time-dependent Ginzburg-Landau equations. In this expression, $L / \xi(T)$ is of the order of ~ 10, $\left( \Delta F / k_B T \right)^{1/2}$ is of the order of ~ 10 and $\left( 1 / \tau_{GL} \right)$ is of the order of ~ $10^{12}$. Hence $\left( 1 / \tau_{GL} \right)$ is the dominant term in the expression for $\Omega$. We also attempt to obtain the estimates of the thermal phase slip rate without relaying on time-dependent Ginzburg-Landau equations, and arrive practically at the same conclusions, as is explained in detail below.

For curve 3, we have replaced $\left( 1 / \tau_{GL} \right)$ by the characteristic frequency of the nanowire, which acts as an inductor and forms an $LC$-circuit with the leads, which are coupled to each other by a capacitance. In other words, we replace $\left( 1 / \tau_{GL} \right)$ with $\omega_0 = 1 / \sqrt{L_w C}$, where $L_w \simeq \hbar L / 3\sqrt{3} e I_C(T) \xi(T)$ is the kinetic inductance of the wire (*12*), and $C$ is the capacitance of the leads. For the calculations $C$ is taken to be 10 fF (*13*). Thus obtained curve (green line in Fig. S3) is very close to the traditional LAMH result (the red curve). In another attempt to verify the approximate validity of the McCumber-Halperin attempt frequency, we replaced $\left(1/\tau_{GL}\right)$ by a well-known expression of plasma frequency for a JJ (*2*), i.e., $\omega_p = \sqrt{2eI_C(T)/\hbar C}$. Again, the obtained curve (grey) appears very close to the LAMH result. Thus, in all cases, we find that with the TAPS model, the prediction of the phase slip rate is ~ $10^{15}$ orders of magnitude smaller than the data and we can in no way account for this difference by changing the attempt frequency. Hence, it strongly indicates that, at low temperatures, the measured phase slips are QPS, not TAPS.

We can also estimate the zero-bias resistance from our high-bias switching current measurements for very low temperatures, by an extrapolation. Using eq. 2, we can convert the zero-bias phase slip rate, $\Omega(T)\exp\left(-\dfrac{\Delta F(T)}{k_B T}\right)$ to resistance. We find that the resistance drops exponentially from $10^{-50}$ to $10^{-80}$ $\Omega$ for temperatures from 1.1 K to 0.3 K in the QPS dominated regime. This resistance is obviously very small to be measured in a typical lab setup and can only be estimated from such an extrapolation of the switching current measurements data. We verified that this resistance is of the same order as predicted by Golubev-Zaikin (GZ) theory (*14*), which gives for zero-temperature limit the result as follows, $R_{QPS} = \Omega(T)\exp(-AR_Q L/R_N\xi(0))$. To get the resistance value of $10^{-50}$ to $10^{-80}$ $\Omega$ we varied $A$ from 2.7 to 4.0 for T = 1.1 K to T = 0.3 K. This is in agreement with the GZ theory, since they predict that $A$ should be of the order of unity, which we confirm.

## Macroscopic quantum tunnelling in high-$T_C$ intrinsic Josephson junctions

In order to verify that we can observe MQT in our [3]He setup, we measured two high-$T$c crystal samples with intrinsic Josephson junctions (IJJ), using the same measurement scheme that was used for the nanowire sample measurements. The general idea in doing this was that the MQT in high-$T$c stacked junctions is well known and well understood. By observing the results seen by other groups we hope to gain extra confidence in the correctness of our setup. Indeed, the results obtained in such test confirm that the setup is working properly, as is explained in more details below.

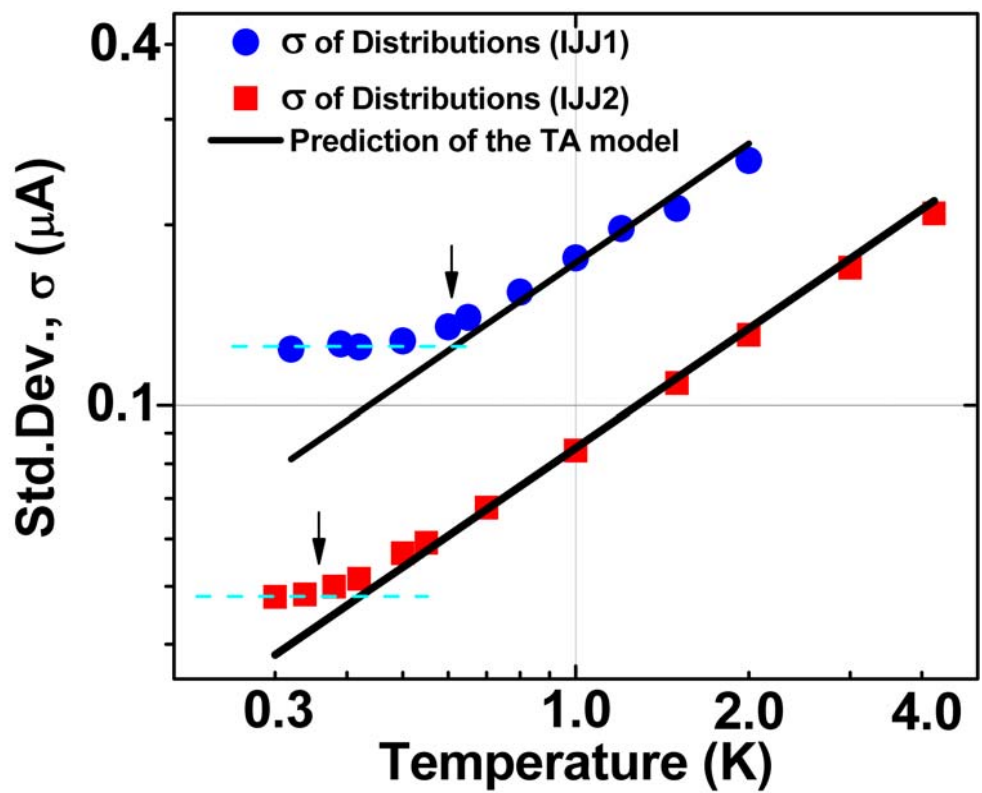

FIG. S4: Standard deviation (σ) of switching current distributions vs. temperature for two high-$T$c crystal with intrinsic Josephson junction samples (IJJ1 and IJJ2), measured in the same [3]He system in which all the nanowire samples were measured.  We clearly see the MQT regime (denoted by the saturation of the distribution width) below a crossover temperature $T^*$ = 0.65 K for IJJ1 and $T^*$ = 0.35 K for IJJ2 (indicated by the two arrows). In the high temperature range, as predicted by the thermal activation model, σ is proportional to $T^{2/3}$ (solid black line). The fluctuation free critical currents for the samples are 170.2 μA (IJJ1) and 17.6 μA (IJJ2).

Fig. S4 shows the standard deviation of the switching current distributions as a function of temperature obtained from the two samples, IJJ1 and IJJ2.  The samples were fabricated from $Bi_2Sr_2CaCu_2O_{8+x}$ crystal shaped using focused ion beam (FIB) to the lateral dimensions of $1.6 \times 2.4$ μm$^2$ (IJJ1).  The bias current in these measurements was injected parallel to the c-axis (i.e., perpendicular to the weakly coupled superconducting planes of the crystal).  In Fig. S4, we observe a crossover from a thermal activated escape regime to MQT regime near $T^*$ = 0.65 K for IJJ1 and near $T^*$ = 0.35 K for IJJ2, which is manifested by a saturation behaviour of the standard deviation at lowering temperatures ([15]).  We also find that, in the high temperature range, σ is proportional to $T^{2/3}$ which is expected for a thermally activated escape model ([16]).  To validate this further, we estimated the escape temperature, $T_{esc}$, from the escape rates, Γ (obtained from switching current distributions) at different temperatures (see Fig. S5 a).  For this, we used the usual expression, $\Gamma = (\omega_p / 2\pi) \exp(-\Delta U / k_B T_{esc})$, where $\omega_p$ is the plasma frequency and $\Delta U = \left(4\sqrt{2} I_0 \Phi_0 / 6\pi\right)\left(1 - I / I_0\right)^{3/2}$ ($I_0$ is the fluctuation free critical current) the barrier energy for escape of the "phase particle"([11]). The obtained $T_{esc}$ is plotted versus the bath temperature, $T_{bath}$ in Fig. S5 b.  We find that, for, high temperatures, $T_{esc} = T_{bath}$, indicating the escape process is dominated by thermal activation. For $T_{bath} < 0.65$ K, $T_{esc}$ saturated to a value of 0.73 K, indicating a region where escape process is dominated by MQT (*11*).

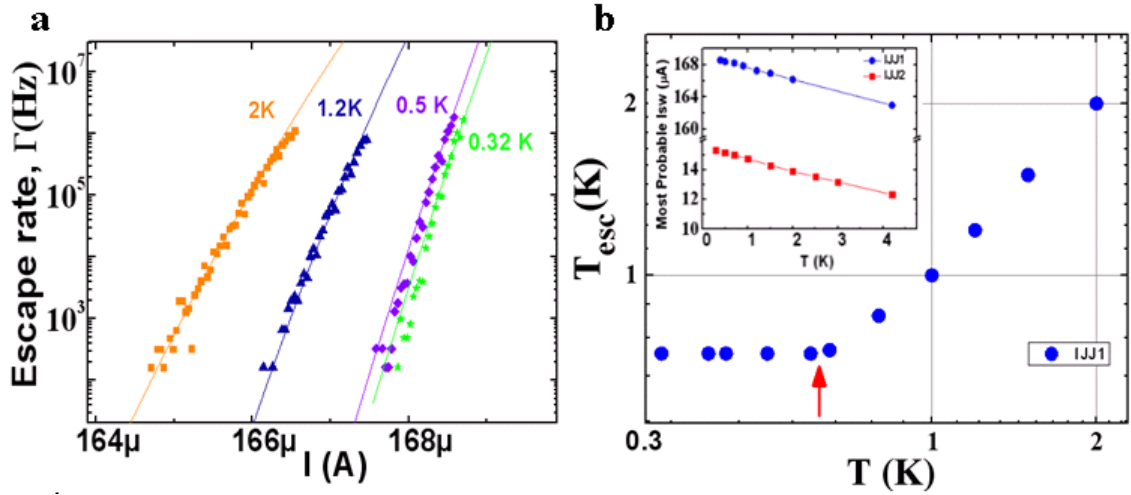

**Fig. S5: a,** Switching rates obtained from the switching current distributions (symbols) and the fits obtained (solid lines) using the expression $\Gamma = (\omega_p / 2\pi) \exp(-\Delta U / k_B T_{esc})$ for sample IJJ1 (see text). **b,** The escape temperature $T_{esc}$ obtained at various bath temperatures. $T_{esc}$ saturates below ~ 0.65 K (indicated by the arrow). The inset shows the most probable switching currents for the two samples obtained from switching current distributions. The critical current for IJJ1 is ~ 10 times larger than IJJ2.

Furthermore, to verify that the saturation in σ is not due to noise (i.e., the electronic temperature not decreasing below 0.6 K), we measured another sample (IJJ2) with a critical current ~ 10 times smaller than IJJ1 ($I_C(0)$ for IJJ1 ~170.2 μA, $I_C(0)$ for IJJ2 ~17.6 μA). As shown in Fig. S4, the σ follows the prediction of the thermal activation to a lower crossover temperature of ~ 0.35 K, as is expected for a sample with a lower critical current (since the crossover temperature is proportional to $\sqrt{I_C / C}$ where $I_C$ is the critical current and $C$ is the junction capacitance) (*15*). Also, for both the samples, the most probable switching current increases with temperature decreasing, indicating that the sample temperature decreases with the bath temperature, down to the lowest attainable temperature (see the inset of Fig. S5 b).

The observation of crossover temperature (i.e., the observation of MQT) in high-$T$c crystals with weakly coupled superconducting planes indicates that the unexpected behaviour in $\sigma(T)$ of a superconducting nanowire is not due to some uncontrolled environmental noise but originates from an intrinsic quantum fluctuations in these samples.

## Filtering system in our measurement setup:

In this section we discuss the arrangement of RF filters in our [3]He measurement setup. The purpose of these filters is to suppress external high-frequency electromagnetic noise, such as the noise originating from cell-phones, radio stations, and also the black-body radiation, which can, if filters are not installed, propagate through the measurement leads and reach the sample and modify the switching current observed in the experiment. Our filters are designed to reduce this noise effect to a negligible level.

Our main filtering stage is a Copper powder filter thermalized at the base temperature (0.29 K). The filter is of the type developed by Martinis, Devoret and Clarke (*11*). More details are presented below.

In our system, each signal line has three stages of filtering in series, namely, a $\pi$-filter at room temperature and a copper-powder filter (Cu-F) (at the base temperature) and silver-paste filter (Ag-F) (also, at the base temperature). These filters are necessary to suppress noise ranging from low frequency to high microwave frequencies. The compact powder filters (i.e., Cu-F or Ag-F) rely on the skin effect damping for attenuation of high frequencies. At room temperature, commercially available π-filters (Spectrum Control, SCI 1201-066) are placed on each electrical lead before they enter the cryostat. The π-filters are mounted inside an aluminium box (Hammond Manufacturing) which is attached to the top of the cryostat. The π-filters used are low-pass filters with a rated 7 dB cut-off frequency of 3 MHz. As shown in Fig. S6, for frequencies larger than 10 MHz, the measured attenuation of these π-filters is more than 20 dB. Our copper powder filters are fabricated using three feet of coiled insulated Constantan wire [Cu(55 %)Ni(45 %) alloy, resistance 18.4 Ω/feet, diameter 0.004 inch] embedded in a mixture of copper powder (-325 mesh, Alfa Aesar) and epoxy (Stycast # 1226, Emerson and Cuming). Similarly, the silver paste filters are fabricated using three feet of coiled insulated Constantan wire (the same wire) in silver paste (Fast drying silver paint, Ted Pella Inc.). By measuring the signal lines with all the filters, using a vector network analyzer (Agilent N5230A), we found the attenuation to be larger that 100 dB for frequencies higher than 1 GHz. Any frequency above 6.25 GHz (which corresponds to a temperature of 0.3 K) is attenuated by more than 110 dB and falls below the noise floor of our network analyzer. This level of attenuation is similar to the attenuation used in previous experiments on MQT, see for example, ref. 11. In addition, the test performed on wires with different critical currents and on high-$T$c samples with different critical currents indicate that MQT becomes dominant at higher temperatures in samples with higher critical currents. This is a good proof of the fact that the observed behaviour is really due to MQT and not due to a noisy environment. For the corresponding discussion see ref. 2, page 262.

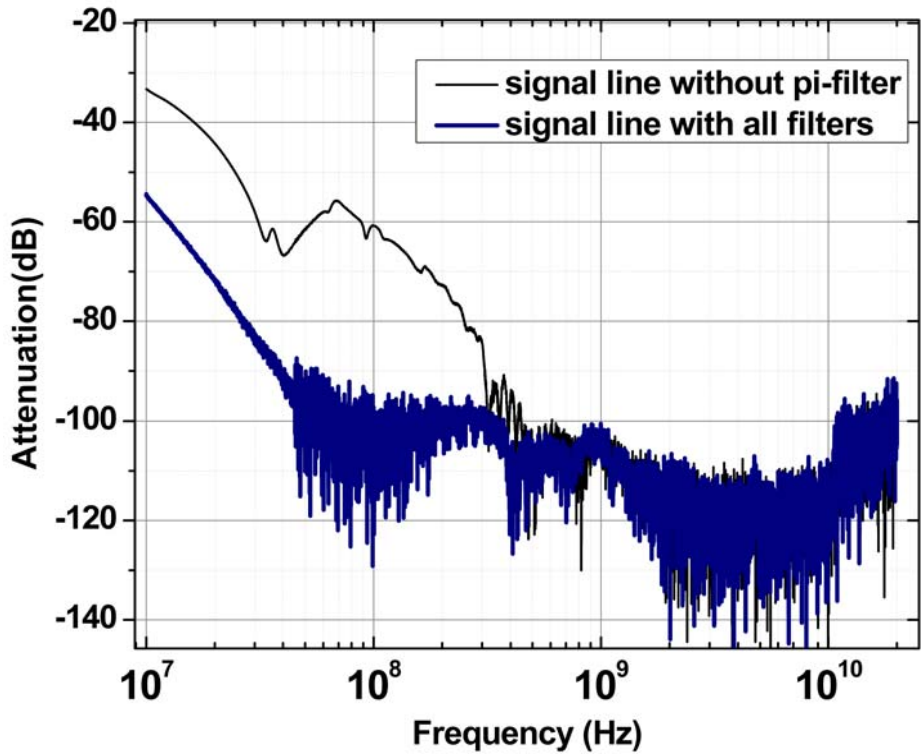

**FIG. S6:** Attenuation of a signal line of our [3]He setup measured at room temperature between 10 MHz and 20 GHz. For measurement of the signal line with all the three stages of filters (blue curve) we find attenuation larger than 90 dB for frequencies higher than 40 MHz. For frequencies higher than 6 GHz (roughly corresponding to our base temperature of $T \sim 0.29$ K), we find the attenuation to be larger than 110 dB and the signal falls below the noise level of our network analyzer. The attenuation of the signal lines without the π-filter is also shown (black curve). The π-filters provide an attenuation of 20 dB for frequencies larger than 10 MHz (rated 7 dB cut-off frequency of 3 MHz).

**Supplementary References:**